\documentclass[12pt]{article}
\usepackage{epsf,amsmath}

\def\qslash{\rlap{/}{q}}

\def\nslash{\rlap{\hspace{0.02cm}/}{n}}

\def\spose#1{\hbox to 0pt{#1\hss}}
\def\lsim{\mathrel{\spose{\lower 3pt\hbox{$\mathchar"218$}}
 \raise 2.0pt\hbox{$\mathchar"13C$}}}
\def\gsim{\mathrel{\spose{\lower 3pt\hbox{$\mathchar"218$}}
 \raise 2.0pt\hbox{$\mathchar"13E$}}}

\begin{document}

\begin{titlepage}

\begin{flushright}
{\small
CLNS~02/1802\\
PITHA~02/14\\
hep-ph/0210085\\
October~4, 2002}
\end{flushright}

\vspace{0.7cm}
\begin{center}
{\Large\bf Flavor-Singlet B-Decay Amplitudes in\\[0.3cm]  
QCD Factorization}
\end{center}

\vspace{0.8cm}
\begin{center}
{\sc Martin Beneke$^a$ and Matthias Neubert$^b$}\\
\vspace{0.7cm}
{\sl ${}^a$Institut f\"ur Theoretische Physik E, RWTH Aachen\\
D--52056 Aachen, Germany\\
\vspace{0.3cm}
${}^b$Newman Laboratory for Elementary-Particle Physics, Cornell 
University\\
Ithaca, NY 14853, USA}
\end{center}

\vspace{0.3cm}
\begin{abstract}
\vspace{0.2cm}\noindent 
Exclusive hadronic $B$-meson decays into two-body final states consisting 
of a light pseudoscalar or vector meson along with an $\eta$ or $\eta'$ 
meson are of great phenomenological interest. Their theoretical analysis 
involves decay mechanisms that are unique to flavor-singlet states, such 
as their coupling to gluons or their ``intrinsic charm'' content. These 
issues are studied systematically in the context of QCD factorization and 
the heavy-quark expansion. Theory can account for the experimental data 
on the $B\to K^{(*)}\eta^{(\prime)}$ branching fractions, albeit within 
large uncertainties.
\end{abstract}

\vfil
\end{titlepage}

\section{Introduction}

The unexpectedly large branching fraction of order $6\cdot 10^{-4}$ for 
the inclusive production of high-momentum $\eta'$ mesons in $B$ decays 
first reported by the CLEO Collaboration in 1997 \cite{Browder:1998yb} 
has triggered some theoretical activity aiming at explaining the 
dynamical origin of this enhancement. The large mass of the $\eta'$ meson 
as compared to other pseudoscalar mesons is one of the clearest 
manifestations of the axial anomaly in QCD. It is therefore natural to 
ask how the anomaly affects the production process of $\eta'$ mesons
\cite{Atwood:1997bn,Hou:1997wy,Fritzsch:1997ps}. Other possible 
enhancement mechanisms (not mentioning those that invoke new fundamental 
interactions) include a large $c\bar c$ content of the $\eta'$ 
\cite{Halperin:1997as,Yuan:1997ts}. Given the theoretical uncertainties 
involved in the calculation of semi-inclusive branching fractions, the 
dynamical details remain unclear except that there appears to be no 
fundamental problem to account for the observed branching fraction.

Recent data on two-body final states containing $\eta'$ (summarized in 
Table~\ref{tab:data}) show that the branching fraction for 
$B^-\to K^-\eta'$ decay is about six times larger than that for the
corresponding $B^-\to K^-\pi^0$ decay, confirming the semi-inclusive
enhancement. The comparison with other two-body final states allows us to 
extract a more detailed pattern. For instance, if $\eta'$ is replaced by 
$\eta$ the corresponding branching fraction is suppressed rather than 
enhanced compared to the $\pi^0$ mode. Furthermore, the enhancement and 
suppression appears to be absent or even reversed when the pseudoscalar 
kaon is replaced by the corresponding vector meson, and with the current 
limited data no particularly striking pattern is visible when the kaons 
are replaced by pions or $\rho$ mesons.

The theoretical description of exclusive two-body decays has also 
improved, since the concept of factorization is now better understood
\cite{BBNS}. The QCD factorization approach, in particular, 
has successfully explained the magnitude of tree and penguin amplitudes 
of final states with pions and kaons \cite{Beneke:2001ev}, while the 
calculation of strong interaction phases remains to be validated by 
experiments. It is therefore clearly interesting to see whether this 
approach can explain the pattern of branching fractions containing 
$\eta^{(\prime)}$ mesons as described above. The new element that has to 
be understood for this purpose is a flavor-singlet amplitude, defined as 
the amplitude for producing a quark--antiquark pair not containing the 
spectator quark in the coherent flavor state $(u\bar u+d\bar d+s\bar s)$ 
or a pair of gluons, where the quarks or gluons have small relative 
transverse momentum and hadronize into an $\eta^{(\prime)}$ meson. 

The analysis of this singlet amplitude, which has not been considered 
systematically in the QCD factorization approach so far, is the main 
purpose of this paper. We will indeed see below that the standard QCD 
factorization formula does not hold, but that a suitable modification 
allows us to obtain the flavor-singlet amplitude at leading order in an 
expansion in $1/m_b$ at the price of introducing one new non-perturbative 
parameter. We shall find that the QCD factorization approach can 
qualitatively  account for the pattern of exclusive branching fractions 
described above, including the large rate for the $K\eta'$ final state, 
but the theoretical uncertainties are rather large. In particular, we 
find that it is the constructive or destructive interference of 
non-singlet penguin amplitudes rather than an enhanced singlet penguin 
amplitude which is responsible for the distinctive pattern of 
strangeness-changing transitions to $\eta^{(\prime)}$ mesons, much along 
the lines envisaged qualitatively by Lipkin \cite{Lipkin:1990us}. The 
improved description of data compared to the naive factorization 
analysis \cite{Ali:1997ex} comes from the radiative enhancement of the 
non-singlet penguin amplitude, which also underlies the sizeable 
branching fractions for $\pi K$ final states. Our findings are in 
contrast to other recent analyses of the $\eta^{(\prime)}$ modes 
\cite{Ahmady:1997fa,Du:1997hs,Yang:2000ce}, where the large branching 
fractions for the $K\eta^{(\prime)}$ final states were attributed due 
to an enhanced flavor-singlet penguin amplitude. Similarly, the singlet 
amplitudes we obtain are smaller than those inferred from 
phenomenological analyses using SU(3) symmetry and certain dynamical 
assumptions about flavor topologies \cite{Gronau:1999hq}. For 
completeness we note that the decays $B\to K\eta^{(\prime)}$ have also 
been analyzed using the perturbative QCD approach \cite{Kou:2001pm}, 
however no singlet-specific mechanisms besides $\eta$--$\eta'$ mixing 
were investigated in this study. 

\begin{table}
\centerline{\parbox{14cm}{\caption{\label{tab:data}
CP-averaged experimental branching ratios (in units of $10^{-6}$) on 
charmless $B$ decays into two-body final states containing $\eta$, 
$\eta'$ or $\pi^0$. Upper limits are at 90\% confidence level.}}}
\vspace{0.1cm}
\begin{center}
\begin{tabular}{|l|c|c|c|}
\hline\hline
Mode & CLEO \cite{CLEO} & BaBar \cite{BaBar} & Belle \cite{Belle} \\
\hline
$B^-\to K^-\eta'$ & $80_{\,-\phantom{1}9}^{\,+10}\pm 7$ & $67\pm 5\pm 5$
 & $77.9_{\,-5.9\,-8.7}^{\,+6.2\,+9.3}$ \\
$\bar B^0\to\bar K^0\eta'$ & $89_{\,-16}^{\,+18}\pm 9$ & $46\pm 6\pm 4$
 & $68.0_{\,-\phantom{1}9.6\,-8.2}^{\,+10.4\,+8.8}$ \\
$B^-\to K^-\eta$ & $<6.9$ & & $5.2_{\,-1.5}^{\,+1.7}$ \\
$\bar B^0\to\bar K^0\eta$ & $<9.3$ & & \\
$B^-\to K^-\pi^0$ & $11.6_{\,-2.7\,-1.3}^{\,+3.0\,+1.4}$
 & $12.8\pm 1.2\pm 1.0$ & $13.0_{\,-2.4}^{\,+2.5}\pm 1.3$ \\
$\bar B^0\to\bar K^0\pi^0$ & $14.6_{\,-5.1\,-3.3}^{\,+5.9\,+2.4}$
 & $10.4\pm 1.5\pm 0.8$ & $8.0_{\,-3.1}^{\,+3.3}\pm 1.6$ \\
\hline
$B^-\to K^{*-}\eta'$ & $<35$ & & \\
$\bar B^0\to\bar K^{*0}\eta'$ & $<24$ & $<13$ & \\
$B^-\to K^{*-}\eta$ & $26.4_{\,-8.2}^{\,+9.6}\pm 3.3$ & $<33.9$
 & $26.5_{\,-7.0}^{\,+7.8}\pm 3.0$ \\
$\bar B^0\to\bar K^{*0}\eta$ & $13.8_{\,-4.6}^{\,+5.5}\pm 1.6$
 & $19.8_{\,-5.6}^{\,+6.5}\pm 1.7$ & $16.5_{\,-4.2}^{\,+4.6}\pm 1.2$ \\
$B^-\to K^{*-}\pi^0$ & $<31$ & & \\
$\bar B^0\to\bar K^{*0}\pi^0$ & $<3.6$ & & \\
\hline\hline
\end{tabular}
\end{center}
\end{table}

\section{\boldmath Implementation of $\eta$--$\eta'$ mixing\unboldmath}
\label{sec:mix}

In the calculation of weak decay amplitudes with an $\eta^{(\prime)}$ 
meson in the final state, we need several matrix elements of local
operators evaluated between the vacuum and $\eta^{(\prime)}$. These are 
the matrix elements of the flavor-diagonal axial-vector and pseudoscalar 
current densities, 
\begin{equation}\label{deffh}
\begin{aligned}
   \langle P(q)|\bar q\gamma^\mu\gamma_5 q|0\rangle 
   &= -\frac{i}{\sqrt2}\,f_P^q\,q^\mu \,, \qquad &
   2 m_q \langle P(q)|\bar q\gamma_5 q|0\rangle
   &= -\frac{i}{\sqrt2}\,h_P^q \,, \\
   \langle P(q)|\bar s\gamma^\mu\gamma_5 s|0\rangle
   &= -i f_P^s\,q^\mu \,, & 
   2 m_s \langle P(q)|\bar s \gamma_5 s|0\rangle
   &= -i h_P^s \,, 
\end{aligned}
\end{equation}
where $q=u$ or $d$. We assume exact isospin symmetry and identify 
$m_q\equiv\frac12(m_u+m_d)$. We also need the anomaly matrix elements
\begin{equation}\label{anomalyme}
   \langle P(q)|\frac{\alpha_s}{4\pi}\,G_{\mu\nu}^A\,
   \widetilde{G}^{A,\mu\nu}|0\rangle = a_P \,,
\end{equation}
where we use the convention
\begin{equation}
   \widetilde{G}^{A,\mu\nu} = -\frac12\,\epsilon^{\mu\nu\alpha\beta}
   G_{\alpha\beta}^A \qquad (\epsilon^{0123}=-1)
\end{equation}
for the dual field-strength tensor. In all cases $P=\eta$ or $\eta'$ 
denotes the physical pseudoscalar meson state. We also need the 
generalization of the local quark operators to the corresponding 
light-cone operators, which define the twist-2 and twist-3 light-cone 
distribution amplitudes. The treatment of mixing for these 
generalizations will be analogous to the case of the local operators. 
Finally, we need the current matrix elements 
$\langle P|\bar q\Gamma b|\bar B\rangle$, which we decompose into 
Lorentz-invariant form factors as for any other pseudoscalar meson.

The equations above define ten non-perturbative parameters $f_P^i$, 
$h_P^i$, and $a_P$, which however are not all independent. Taking the 
divergence of the flavor-diagonal axial-vector current,
\begin{equation}
   \partial_\mu(\bar q\gamma^\mu\gamma_5 q) = 2im_q\,\bar q\gamma_5 q
   - \frac{\alpha_s}{4\pi}\,G_{\mu\nu}^A\,\widetilde{G}^{A,\mu\nu}
\end{equation}
(and similarly with $q$ replaced by $s$) yields four relations between 
the various parameters, which can be summarized as
\begin{equation}\label{div}
   a_P = \frac{h_P^q-f_P^q\,m_P^2}{\sqrt2} = h_P^s-f_P^s\,m_P^2 \,.
\end{equation}
Without further assumptions this leaves us with six independent 
parameters. It is conventional to write each set of two parameters 
corresponding to $P=\eta,\eta'$, such as $\{f_\eta^s,f_{\eta'}^s\}$, in 
terms of a parameter (such as $f_s$) and a mixing angle (such as 
$\phi_s$). A general treatment then implies three such parameters and 
three independent  mixing angles.
 
In the SU(3) flavor-symmetry limit, where $|\eta\rangle$ is a 
flavor-octet and $|\eta'\rangle$ a flavor-singlet, it follows that 
$f_\eta^s=-\sqrt2\,f_\eta^q$ and $f_{\eta'}^s=f_{\eta'}^q/\sqrt2$, 
$h_\eta^s=-\sqrt2\,h_\eta^q$ and $h_{\eta'}^s=h_{\eta'}^q/\sqrt2$, and 
$a_\eta=0$. However, it is known empirically that SU(3)-breaking 
corrections to these relations are large. In the following we shall not 
rely on SU(3) flavor symmetry but instead introduce another assumption, 
expected to be accurate at the 10\% level, to reduce the number of 
hadronic parameters. This assumption leads to what will be referred to as 
the Feldmann--Kroll--Stech (FKS) mixing scheme \cite{Feldmann:1998vh}. In 
the absence of the axial U(1) anomaly, the flavor states 
$|\eta_q\rangle=(|u\bar u\rangle+|d\bar d\rangle)/\sqrt2$ and 
$|\eta_s\rangle=|s\bar s\rangle$ mix only through OZI-violating effects 
known phenomenologically to be small. We therefore assume that the 
anomaly is the only effect that mixes the two flavor states. (This 
assumption implies, in particular, that the vector mesons $\omega$ and 
$\phi$ are pure $(u\bar u+d\bar d)$ and $s\bar s$ states, respectively, 
as is indeed the case to very good approximation.) In a chiral Lagrangian 
treatment of the pseudoscalar mesons (including the $\eta'$ meson, which 
can be done in a combined chiral and $1/N_c$ expansion 
\cite{Kaiser:1998ds}), the anomaly introduces an effective mass term for 
the system of $\eta^{(\prime)}$ states that is not diagonal in the flavor 
basis $\{|\eta_q\rangle,|\eta_s\rangle\}$, and since this is by 
assumption the only mixing effect, the FKS scheme amounts to a scheme 
with a single mixing angle in the flavor basis. If the physical states 
are related to the flavor states by
\begin{equation}
   \left( \begin{array}{c}
    |\eta\rangle \\ |\eta'\rangle
   \end{array} \right) 
   = \left( \begin{array}{cc}
    \cos\phi & ~-\sin\phi \\  
    \sin\phi & \phantom{~-}\cos\phi
   \end{array} \right) 
   \left( \begin{array}{c}
    |\eta_q\rangle \\ |\eta_s\rangle
   \end{array} \right) ,
\end{equation}
then the same mixing angle applies to the decay constants $f_P^i$ and 
$h_P^i$ with the normalization given by (\ref{deffh}). We therefore write
\begin{equation}
\begin{aligned}
   f_\eta^q &= f_q\cos\phi \,, \qquad & 
   f_\eta^s &= -f_s\sin\phi \,, \\
   f_{\eta'}^q &= f_q\sin\phi \,, &
   f_{\eta'}^s &= f_s\cos\phi \,,
\end{aligned}
\end{equation}
and an analogous set of equations for the $h_P^i$. This defines four new 
parameters $f_{q,s}$ and $h_{q,s}$. Inserting these results into 
(\ref{div}) allows us to express all ten non-perturbative parameters in 
terms of the decay constants $f_q$, $f_s$ and the mixing angle $\phi$. We 
obtain
\begin{equation}\label{hqres}
\begin{aligned}
   h_q &= f_q\,(m_\eta^2\,\cos^2\phi + m_{\eta'}^2\,\sin^2\phi)
    - \sqrt2 f_s\,(m_{\eta'}^2-m_\eta^2)\,\sin\phi\cos\phi \,, \\
   h_s &= f_s\,(m_{\eta'}^2\,\cos^2\phi + m_\eta^2\,\sin^2\phi)
    - \frac{f_q}{\sqrt2}\,(m_{\eta'}^2-m_\eta^2)\,\sin\phi\cos\phi \,,
\end{aligned}
\end{equation}
and
\begin{equation}\label{theas}
\begin{aligned}
   a_\eta &= -\frac{1}{\sqrt2}\,(f_q\,m_\eta^2-h_q)\,\cos\phi
    = -\frac{m_{\eta'}^2-m_\eta^2}{\sqrt2}\,\sin\phi\cos\phi\,
    (-f_q\,\sin\phi+\sqrt2 f_s\,\cos\phi) \,, \\
  a_{\eta'} &= -\frac{1}{\sqrt2}\,(f_q\,m_{\eta'}^2-h_q)\,\sin\phi
   = -\frac{m_{\eta'}^2-m_\eta^2}{\sqrt2}\,\sin\phi\cos\phi\,
   (f_q\,\cos\phi+\sqrt2 f_s\,\sin\phi) \,.
\end{aligned}
\end{equation}

{\arraycolsep0.2cm
\begin{table}[t]
\centerline{\parbox{14cm}{\caption{\label{tab:FKS}
Decay constants (in MeV), pseudoscalar densities and anomaly matrix 
elements (in GeV$^3$) of $\eta^{(\prime)}$ in the FKS mixing scheme.}}}
\vspace{0.1cm}
\begin{center}
$
\begin{array}{|c|r||c|r||c|r|}
\hline\hline
h_q & 0.0015 \pm 0.004 & f_\eta^q & 108\pm 3 & 
h_\eta^q  & 0.001 \pm 0.003 \\
h_s & 0.087 \pm 0.006 & f_\eta^s & -111\pm 6& 
h_\eta^s  & -0.055 \pm 0.003 \\
a_\eta & -0.022 \pm 0.002 & f_{\eta'}^q & 89 \pm 3 & 
h_{\eta'}^q  & 0.001 \pm 0.002 \\
a_{\eta'} & -0.057 \pm 0.002 & f_{\eta'}^s & 136 \pm 6 & 
h_{\eta'}^s  & 0.068 \pm 0.005\\[-0.55cm]
&&&&& \\
\hline\hline
\end{array}
$
\end{center}
\end{table}}

The three remaining parameters of the FKS scheme have been determined 
from a fit to experimental data, yielding \cite{Feldmann:1998vh}
\begin{equation}
   f_q = (1.07\pm 0.02)\,f_\pi, \qquad
   f_s = (1.34\pm 0.06)\,f_\pi, \qquad
   \phi = 39.3^\circ\pm 1.0^\circ,
\end{equation}
where the errors do not include a possible systematic uncertainty from 
the theoretical assumptions underlying the FKS scheme. The numerical 
values of the other parameters derived from these inputs are given in 
Table~\ref{tab:FKS}. As can be seen from the table the expression for 
$h_q$ in (\ref{hqres}) is not useful to compute this parameter accurately 
in practice, since large cancellations occur between the terms 
proportional to $f_q$ and $f_s$. In fact, from (\ref{deffh}) it follows 
that $h_q$ must vanish in the limit $m_q\to 0$ and so is expected to be 
very small ($h_q/h_s\approx m_q/m_s$).\footnote{Because the masses of 
$\eta$ and $\eta'$ are themselves functions of the mixing angle $\phi$, 
it would be inconsistent to interpret (\ref{hqres}) as equations for 
$h_q$ and $h_s$ as functions of $\phi$ with $m_\eta$ and $m_{\eta'}$ 
fixed to their physical values. Rather, only those values of $m_\eta$, 
$m_{\eta'}$ and $\phi$ leading to a small value of $h_q$ are acceptable. 
Setting $h_q=0$ gives $\phi\simeq 38^\circ$, which is close to the 
phenomenological value quoted above.} 
An alternative expression is obtained with the help of the chiral and 
large-$N_c$ expansion. To leading order this gives \cite{Kaiser:1998ds}
\begin{equation}
\begin{aligned}
   f_q &= f_\pi \,, &
   f_s &= \sqrt{2 f_K^2-f_\pi^2} = 1.41 f_\pi \,, \\
   h_q &= f_q\,m_\pi^2 = 0.0025\,\text{GeV}^3 \,, \qquad & 
   h_s &= f_s\,(2m_K^2-m_\pi^2) = 0.086\,\text{GeV}^3 \,.
\end{aligned}
\end{equation}
These results are also prone to large errors, since the chiral expansion 
is expected to apply only if the masses of the pseudoscalar mesons (now 
including the $\eta^\prime$) are much smaller than the masses of the 
other hadrons. We conclude that $h_q$ is poorly determined at present. 
Note that although $h_q$ is small it cannot be neglected in general, 
since some decay amplitudes are proportional to $h_q/m_q$.

\section{The flavor-singlet amplitude}

A distinctive feature of hadronic $B$ decays with an $\eta^{(\prime)}$ 
meson in the final state is that these states contain a flavor-singlet 
component in their wave function, which opens possibilities for novel
decay mechanisms that do not occur if the final state contains only 
flavor non-singlet states such as pions and kaons. In this section we 
discuss the different contributions to the flavor-singlet amplitude in 
the QCD factorization approach. 

Neglecting weak annihilation terms for now, we write the 
$\bar B\to\bar K P$ decay amplitudes (where $P=\eta^{(\prime)}$) as 
\begin{equation}\label{silly}
   {\cal A}(\bar B\to\bar K P)
   = i\,\frac{G_F}{\sqrt2} \sum_{p=u,c} \lambda_p^{(s)}\,{\cal A}_p(K P)
   \,,
\end{equation}
where $\lambda_p^{(s)}=V_{pb} V_{ps}^*$, and 
\begin{eqnarray}\label{PKamp}
   {\cal A}_p(K P) &=& m_B^2\,F_0^{B\to P}(0)\,\frac{f_K}{\sqrt2}
    \left\{ \delta_{pu}\,\delta_{q_s u}\,\alpha_1(P K)
    + \alpha_4^p(P K) + \frac32\,e_{q_s} \alpha_{4,{\rm EW}}^p(P K)
    \right\} \nonumber\\
   &+& m_B^2\,F_0^{B\to K}(0)\, 
    \bigg\{ \frac{f_P^q}{\sqrt2}\,\Big[ \delta_{pu}\,\alpha_2(K P_q)
    + 2\alpha_3^p(K P_q) + \frac12\,\alpha_{3,{\rm EW}}^p(K P_q) \Big]
    \nonumber\\ 
   &&\hspace{2.43cm}\mbox{}+ f_P^s\,\Big[ \alpha_3^p(K P_s)
    + \alpha_4^p(K P_s) - \frac12\,\alpha_{3,{\rm EW}}^p(K P_s)
    - \frac12\,\alpha_{4,{\rm EW}}^p(K P_s) \Big] \nonumber\\
   &&\hspace{2.43cm}\mbox{}+ f_P^c\,\big[ \delta_{pc}\,\alpha_2(K P_c)
    + \alpha_3^p(K P_c) \big] \bigg\} \,.
\end{eqnarray}
Here $F_0^{B\to M}(0)$ is a $B\to M$ transition form factor evaluated at 
$q^2=0$, $q_s=u$ or $d$ denotes the flavor of the spectator antiquark in 
the $\bar B$ meson, $e_u=2/3$ and $e_d=-1/3$ are the corresponding 
electric charge factors, and the symbol $\delta_{q_s u}$ (and similarly 
$\delta_{p u}$) is defined such that $\delta_{q_s u}=1$ if $q_s=u$ and 
zero otherwise. The notation $P_i$ will be explained below. 

The various $\alpha_i$ coefficients in the expression for the amplitudes 
correspond to different flavor topologies contributing to a given decay 
\cite{inprep}. The arguments in parentheses are such that the first meson 
inherits the spectator antiquark from the $B$ meson, while the second 
meson is the ``emission particle'' produced at the weak vertex. In naive 
factorization, the $\alpha_i$ are given in terms of combinations of 
Wilson coefficients of the effective weak Hamiltonian as
\begin{equation}\label{NF}
\begin{aligned}
   \alpha_1(P K) &= C_1 + \frac{C_2}{N_c} \,, \qquad &
   \alpha_4^p(P K) &= C_4 + \frac{C_3}{N_c}
    + r_\chi^K \left( C_6 + \frac{C_5}{N_c} \right) , \\
   \alpha_2(K P_i) &= C_2 + \frac{C_1}{N_c} \,, &
   \alpha_4^p(K P_s) &= C_4 + \frac{C_3}{N_c}
    + r_\chi^{P_s} \left( C_6 + \frac{C_5}{N_c} \right) ,
\end{aligned}
\end{equation}
and
\begin{equation}\label{al3NF}
   \alpha_3^p(K P_i) = C_3 + \frac{C_4}{N_c}
   - \left( C_5 + \frac{C_6}{N_c} \right) ,
\end{equation}
where $r_\chi^K=2 m_K^2/[m_b (m_s+m_q)]$ in the definition of 
$\alpha_4^p(P K)$, and $r_\chi^{P_s}=h_s^P/(f_s^P m_b m_s)$ in the case 
of $\alpha_4^p(K P_s)$. Analogous expressions hold for the electroweak 
penguin coefficients $\alpha_{3,{\rm EW}}^p$ and $\alpha_{4,{\rm EW}}^p$.
Note that the singlet amplitude $\alpha_3^p(K P_i)$ is generated by 
penguin operators in the effective weak Hamiltonian through transitions 
of the type $b\to s q\bar q$, where the $q\bar q$-pair hadronizes into an 
$\eta^{(\prime)}$ meson.

The calculation of corrections of order $\alpha_s$ to the coefficients 
$\alpha_{1,2,4}$ is analogous to the calculation for pions or kaons 
discussed in \cite{Beneke:2001ev}. (In the notation of that paper, we 
have $\alpha_{1,2}=a_{1,2}$, $\alpha_3=a_3-a_5$, 
$\alpha_4=a_4+r_\chi a_6$, $\alpha_{3,{\rm EW}}=a_9-a_7$, and 
$\alpha_{4,{\rm EW}}=a_{10}+r_\chi a_8$.) In the FKS scheme, the 
$\eta^{(\prime)}$ system is described by two leading-twist 
quark--antiquark light-cone distribution amplitudes, $\phi_q(x)$ and 
$\phi_s(x)$, defined in analogy with the decay constants $f_q$ and $f_s$. 
The subscript on $P_q$ and $P_s$ in (\ref{PKamp}) means that the 
corresponding distribution amplitude must be used in the expressions for 
the $\alpha_i$ parameters. The one-loop vertex corrections to the 
quark--antiquark amplitude and a contribution from spectator scattering 
to the singlet amplitude have also already been computed in 
\cite{Beneke:2001ev}. This result is not complete, however, since it 
ignored the possibility that the $\eta^{(\prime)}$ meson is formed from 
two gluons. 

In the following subsections we discuss three contributions to the 
singlet decay amplitude related to the gluon content of the 
$\eta^{(\prime)}$: the $b\to sgg$ amplitude and its relation to an 
effective charm decay constant, spectator scattering involving two 
gluons, and singlet weak annihilation, the latter being suppressed by at 
least one power of $\Lambda/m_b$ in the heavy-quark limit. Before we 
begin this discussion we comment on the treatment of the heavy-quark 
content of the $\eta^{(\prime)}$ meson. In QCD factorization the decay 
amplitude is factorized into short-distance kernels and light-cone 
distribution amplitudes. When the factorization scale is smaller than the 
heavy-quark mass, heavy-quark contributions are part of the 
short-distance kernels, and the light meson is described in terms of 
distribution amplitudes of light quarks and gluons. If, in the limit 
$m_b\to\infty$, the charm-quark mass is held fixed, one should still 
introduce charm light-cone distribution amplitudes to sum large 
logarithms of $m_b/m_c$, but these distribution functions (and the 
associated decay constants) can be computed in terms of those of light 
quarks. In (\ref{PKamp}) we included the possibility of such an induced 
charm contribution in the form of terms $f_c^P\alpha_i(K P_c)$, to which 
we will return below.

\subsection{The \boldmath$b\to s gg$ amplitude\unboldmath}

\begin{figure}
\epsfxsize=3.5cm
\centerline{\epsffile{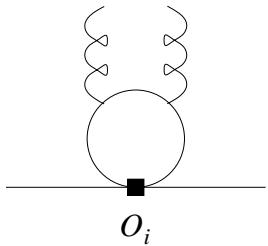}}
\vspace{0.0cm}
\centerline{\parbox{14cm}{\caption{\label{fig:2gluon_kernel}
Two-gluon emission from a quark loop. A second diagram with the two 
gluons crossed is implied.}}}
\end{figure}

\begin{figure}
\epsfxsize=14.0cm
\centerline{\epsffile{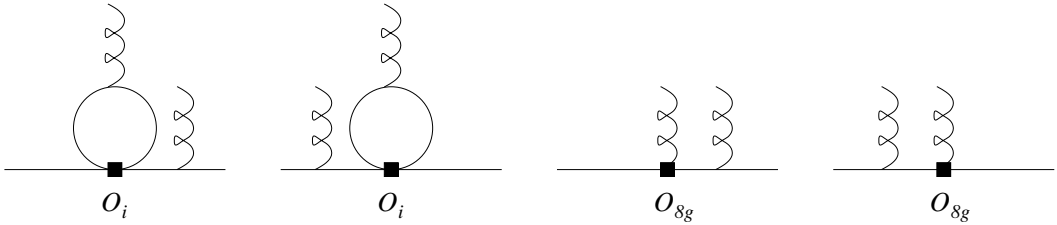}}
\vspace{0.0cm}
\centerline{\parbox{14cm}{\caption{\label{fig:other_2gluon}
Contributions to the $b\to s gg$ amplitude that are power suppressed in 
the heavy-quark limit.}}}
\end{figure}

The $b\to s gg$ amplitude for general gluon momenta has been calculated 
in \cite{Simma:nr} in the electroweak theory. What we need here is the 
special case where the two gluons have small invariant mass to form an 
$\eta^{(\prime)}$, and where the $b\to s$ transition is induced by an 
operator in the effective weak Hamiltonian. The relevant diagrams are 
shown in Figures~\ref{fig:2gluon_kernel} and \ref{fig:other_2gluon}. 

The diagram of Figure~\ref{fig:2gluon_kernel} is only relevant when the 
internal quark is a charm or bottom quark. (Top quarks have already been 
removed from the effective weak Hamiltonian.) Diagrams with light-quark 
loops are governed by hadronic scales and their short-distance part 
consists of the $O_i$ insertion only. This is already taken into account 
in the naive factorization results (\ref{NF}) and (\ref{al3NF}). For an 
operator $O=(\bar s\Gamma_1 b)(\bar q\Gamma_2 q)$, where the $\Gamma_i$ 
denote arbitrary spinor and color matrices, we find (assuming the two 
gluons to be in a color-singlet configuration)
\begin{equation}
   {\cal A}(b\to s g g)|_{{\rm Fig.~\ref{fig:2gluon_kernel}}} 
   = - (\bar u_s\Gamma_1 u_b)\,
   \langle g(q_1) g(q_2)|\text{tr}\,(\Gamma_2\,A)|0\rangle \,, 
\end{equation}
where $u_i$ denote quark spinors, and the trace is performed over spin 
and color indices. The quantity $A$, which is proportional to the 
$3\times 3$ identity matrix in color space, is given by 
\begin{eqnarray}\label{qloop}
   A &=& \frac{\alpha_s}{4\pi N_c}\,\Bigg\{
   \frac{1}{8 m_q}\,\Big[ (1-4r)\,F(r) + 4r \Big]\,
    G_{\mu\nu}^A\,G^{A,\mu\nu}
    + \frac{1}{8 m_q}\,F(r)\,i\gamma_5\,
    G_{\mu\nu}^A\,\widetilde{G}^{A,\mu\nu} \nonumber\\ 
   &&\hspace{3.0cm}\mbox{}+ \frac{1}{q^2}\,\Big[ 1 - F(r) \Big]\,
   iq^\alpha\gamma_\beta\gamma_5\,
   G_{\mu\alpha}^A\,\widetilde{G}^{A,\mu\beta} \Bigg\} \,,
\end{eqnarray}
where $m_q$ is the quark mass in the loop, $r=m_q^2/q^2-i\epsilon$, and 
$q^2=(q_1+q_2)^2$ will later be identified with the square of the 
pseudoscalar meson mass. The function
\begin{equation}
   F(r) = 4r\arctan^2\!\left( \frac{1}{\sqrt{4r-1}} \right)
   = 1 + \frac{1}{12r} + O(1/r^2)
\end{equation}
describes the dependence on the invariant mass $q^2$ of the two gluons 
and the quark mass in the loop. The last term in (\ref{qloop}) may be 
simplified using the identity
\begin{equation}
   G_{\mu\alpha}^A\,\widetilde{G}^{A,\mu}{}_\beta
   = \frac{g_{\alpha\beta}}{4}\,G_{\mu\nu}^A\,\widetilde{G}^{A,\mu\nu}
   \,,
\end{equation}
which for local operators is implied by the antisymmetry of the 
field-strength tensor. In the limit where $m_q^2\gg q^2$ (i.e., for 
$q=c,b$), we obtain
\begin{equation}
   A = \frac{\alpha_s}{4\pi N_c}\,\bigg\{
   \frac{1}{12 m_q}\,G_{\mu\nu}^A\,G^{A,\mu\nu}
   - \frac{(\qslash-6 m_q)\,i\gamma_5}{48 m_q^2}\,
   G_{\mu\nu}^A\,\widetilde{G}^{A,\mu\nu} + O(1/m_q^3) \bigg\} \,.
\end{equation}

For a pseudoscalar state only the matrix element of 
$G_{\mu\nu}^A\,\widetilde{G}^{A,\mu\nu}$ is non-zero. Evaluating the 
Dirac and color structures for the various operators in the effective 
weak Hamiltonian, we obtain for the contribution to the 
$\bar B\to \bar K P$ decay amplitudes
\begin{equation}\label{anomaly}
   {\cal A}_p^{\rm charm} = \frac{a_P}{12 m_c^2}\,
   \langle \bar K(p')|\bar s\qslash(1-\gamma_5)b|\bar B(p)\rangle 
   \left\{ \left( C_2 + \frac{C_1}{N_c} \right)
   \delta_{pc}+ \left( C_3-C_5 + \frac{C_4-C_6}{N_c} \right)
   \right\} ,
\end{equation}
where $q=p-p'$ and where we have neglected the very small contribution
from bottom-quark loops. The definition (\ref{anomalyme}) has been used 
to evaluate the matrix elements of the anomaly. It follows from 
(\ref{theas}) that, to a very good approximation (i.e., neglecting the 
small quantity $h_q$), $a_P/m_P^2=-f_P^q/\sqrt2$. Comparing 
(\ref{anomaly}) with (\ref{PKamp})--(\ref{al3NF}), we see that the 
charm-loop contribution can be incorporated into (\ref{PKamp}) if we 
identify 
\begin{equation}\label{cdec}
   f_P^c = \frac{a_P}{12 m_c^2} 
   \approx -\frac{m_P^2}{12 m_c^2}\,\frac{f_P^q}{\sqrt 2}
\end{equation}
with the leading-order result (in an expansion in $1/m_c$) for a new 
decay constant defined as
\begin{equation}
   \langle P(q)|\bar c\gamma^\mu\gamma_5 c|0\rangle
   = -i f_P^c\,q^\mu \,,
\end{equation}
which might be interpreted as a measure of the ``intrinsic charm'' 
component of the $\eta^{(\prime)}$ wave function. With our definitions 
$f_P^c$ is negative and takes values $f_\eta^c\approx -1$\,MeV and 
$f_{\eta'}^c\approx -3$\,MeV. These results for the decay constants 
agree with a different derivation in \cite{Franz:2000ee}. A similar 
discussion of the charm-loop diagram in the context of non-leptonic $B$ 
decays has already been given in \cite{Ali:1997ex}.

Note that the charm decay constant is formally power-suppressed for heavy
charm quarks (but not power suppressed if only $m_b$ goes to infinity), 
but that the factor of $\alpha_s$ has disappeared in the evaluation of 
the anomaly matrix element. The charm contributions are potentially 
non-negligible, since they multiply the large Wilson coefficients 
$C_{1,2}$. The eventual smallness of this contribution is due to the 
factor $1/12$ in (\ref{cdec}), and because the large Wilson coefficients 
enter in the color-suppressed combination $C_2+C_1/N_c$. Since we have 
identified a new leading-order contribution to the factorized amplitude 
(\ref{PKamp}), we should in principle consider $\alpha_s$ corrections to 
$\alpha_{2,3}(KP_c)$, which would require the calculation of two-loop 
diagrams. We expect that these diagrams can be factored into genuine 
$\alpha_s^2$ corrections and a contribution with the structure of a 
one-loop kernel (with external charm quarks) folded with a calculable 
charm-quark distribution amplitude. Due to the smallness of the 
leading-order charm contributions, however, the calculation of this 
correction is not required within the accuracy of our analysis. 

There are other diagrams that could potentially contribute to the 
$b\to s gg$ amplitude, in which one or both gluons are emitted from 
external quark lines. Figure~\ref{fig:other_2gluon} shows the relevant 
graphs with one gluon emitted from a penguin loop or from the 
chromo-magnetic penguin operator, and the other gluon emitted from the 
$b$ or $s$-quark lines. The loop diagrams are non-zero only for an 
insertion of the penguin operator $O_5$. In order to obtain a 
leading-power contribution to the decay amplitude we must pick out the 
large components of all momenta in the above expression. However, setting 
$q_1=x q+\dots$, $q_2=\bar x q+\dots$ and $p_b-p_s=q+\dots$ one finds 
that the leading term is proportional to $q^\alpha$ or $q^\beta$ and thus 
vanishes for on-shell gluons. It follows that the diagrams in 
Figure~\ref{fig:other_2gluon} are power suppressed with respect to the 
diagram in Figure~\ref{fig:2gluon_kernel}. A similar analysis shows that 
also the graphs where both gluons are emitted from the $b$ and $s$-quark 
lines yield a power-suppressed contribution. 

\subsection{Spectator mechanism}
\label{sec:spec}

\begin{figure}
\epsfxsize=10.0cm
\centerline{\epsffile{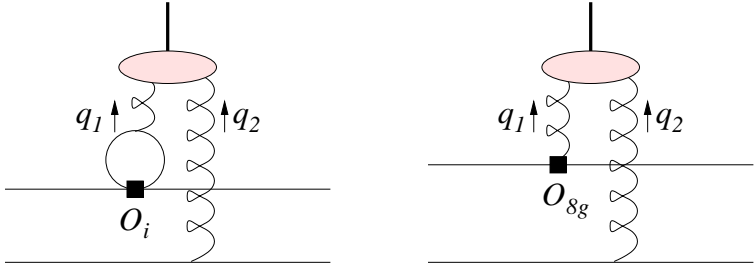}}
\vspace{0.2cm}
\centerline{\parbox{14cm}{\caption{\label{fig:2gluon_O8}
Spectator-scattering contributions to the $B\to K^{(*)}\eta^{(\prime)}$ 
decay amplitudes. The shaded blob represents the 
$\eta^{(\prime)} g^* g^*$ form factor.}}}
\end{figure}

The mechanisms described so far in this section provide the leading
contributions to the flavor-singlet amplitude in the heavy-quark limit,
which start at zeroth order in $\alpha_s$ (naive factorization). In 
practice, these contributions are numerically rather small, because they 
are suppressed by the small Wilson coefficients $C_{3\dots 6}$ or the 
mass ratio $m_P^2/(12 m_c^2)$. We will now identify another leading-power 
contribution to the flavor-singlet amplitude, which starts at order 
$\alpha_s$ but is free of such suppression factors. This contribution is 
related to the spectator-scattering diagrams shown in 
Figure~\ref{fig:2gluon_O8}. 

The flavor-singlet spectator mechanism was considered first in 
\cite{Ahmady:1997fa,Du:1997hs,Yang:2000ce} using the perturbative QCD 
approach. In fact, the kinematics of the spectator graph is such that at 
leading power the two gluons cannot be collinear with the outgoing 
$\eta^{(\prime)}$ meson, and at least one of them must be a (semi-) hard 
gluon. Specifically, assigning momentum 
$p_{\bar q}=\bar y p_K+k_\perp+\dots$ to the antiquark in the kaon (with 
$\bar y=1-y$) and a soft momentum $l$ to the spectator quark in the $B$ 
meson, it follows that for generic values of $y$ not close to 0 or 1 the 
virtualities of the two gluons are 
$q_2^2\approx -2\bar y\,p_K\cdot l=O(m_b\Lambda)$ and 
$q_1^2\approx\bar y m_B^2=O(m_b^2)$. The coupling of two hard gluons to 
the $\eta^{(\prime)}$ meson can be described by a perturbative form 
factor $F(q_1^2,q_2^2)=O(\alpha_s)$ defined in terms of the 
$\eta^{(\prime)} g^* g^*$ vertex as \cite{Muta:1999tc,Ali:2000ci}
\begin{equation}
   i F_P(q_1^2,q_2^2)\,\epsilon_{\mu\nu\alpha\beta}\,
   q_1^\mu\,q_2^\nu\,\delta_{AB}\,\varepsilon_A^{\alpha*}(q_1)\,
   \varepsilon_B^{\beta*}(q_2) \,,
\end{equation}
where $P=\eta'$ or $\eta$. It follows that for generic values of $y$ the 
hard spectator-scattering contribution is an effect of order $\alpha_s^2$ 
and thus is beyond the accuracy of a next-to-leading order analysis. 

Nevertheless, it is instructive to work out the result for this 
contribution. In the kinematic region where $|q_1^2|\gg|q_2^2|$ the 
leading contribution to the form factor can be written as
\begin{equation}\label{ffvalue}
   F_P(q_1^2,q_2^2) = \frac{3 g_s^2}{N_c\,q_1^2}\,C_P \,, \qquad
   (|q_1^2|\gg|q_2^2|)
\end{equation}
where
\begin{equation}\label{CPdef}
   C_P = \sqrt2 f_P^q \int_0^1\!dx\,\frac{\phi_q(x)}{6x\bar x} 
   + f_P^s \int_0^1\!dx\,\frac{\phi_s(x)}{6x\bar x} 
   + \dots
\end{equation}
contains convolution integrals involving the leading-twist 
quark--antiquark distribution amplitudes of the $\eta^{(\prime)}$ meson 
mentioned earlier (twist-3 quark--antiquark distribution amplitudes do 
not contribute due to chirality conservation), and the dots represent a 
contribution involving the two-gluon distribution amplitude, which we 
omit for simplicity. In the limit of asymptotic distribution amplitudes 
this latter contribution vanishes, and $C_P\to\sqrt2 f_P^q+f_P^s$. At 
leading order in $1/m_b$, we find for the hard spectator-scattering 
contribution to the $B\to K\eta^{(\prime)}$ decay amplitude
\begin{equation}\label{hard}
   {\cal A}_p^{\rm spec} = \frac{3C_F\alpha_s^2}{4 N_c^2}\,C_P\,
   f_B f_K\,\frac{m_B}{\lambda_B} 
   \int_0^1 \frac{dy}{\bar y} \left[ P_2^p(y)\,\phi_K(y)
   + r_\chi^K\,P_3^p(y)\,\phi_p^K(y) \right] ,
\end{equation}
where $\phi_K(y)\approx 6y\bar y$ is the leading-twist distribution 
amplitude of the kaon, $\phi_p^K(y)\approx 1$ is one of the twist-3 
amplitudes, and the hadronic parameter $\lambda_B$ is defined as the 
first inverse moment of one of the two leading $B$-meson distribution 
amplitudes \cite{BBNS,Beneke:2001ev}. The quantities $P_{2,3}^p(y)$ are 
the penguin kernels introduced in (49) and (54) of \cite{Beneke:2001ev}.
Explicitly, we have 
\begin{eqnarray}\label{P2y}
   P_2^p(y) &=& C_1 \left[ \frac43\ln\frac{m_b}{\mu} + \frac23
    - G(s_p,\bar y) \right]
    + C_3 \left[ \frac83\ln\frac{m_b}{\mu} + \frac43
    - G(0,\bar y) - G(1,\bar y) \right] \nonumber\\
   &+& (C_4+C_6) \left[ \frac{4n_f}{3}\ln\frac{m_b}{\mu}
    - (n_f-2)\,G(0,\bar y) - G(s_c,\bar y) - G(1,\bar y) \right]
    - \frac{2 C_{8g}^{\rm eff}}{\bar y} \,, \qquad
\end{eqnarray}
with $s_q=(m_q/m_b)^2$, $n_f=5$, and the penguin function $G(s,x)$ as
defined in the same reference. The expression for $P_3^p(y)$ is identical 
to that for $P_2^p(y)$, except that the term  $-2C_{8g}^{\rm eff}/\bar y$ 
must be replaced with $-2C_{8g}^{\rm eff}$. Inspection of the above 
expressions shows that the perturbative hard-scattering contribution has 
a logarithmic endpoint singularity as $y\to 1$, corresponding to the 
region where the gluon exchanged with the spectator quark becomes a soft 
gluon. For the contribution proportional to $C_{8g}^{\rm eff}$ both the 
leading and subleading-twist terms are divergent, whereas for the penguin 
loop contributions only the twist-3 term diverges. In 
\cite{Du:1997hs,Yang:2000ce} these singularities were regularized by 
keeping subleading terms in the quark propagators, which effectively 
corresponds to adopting a model for the non-perturbative endpoint region. 
While this may be legitimate for obtaining a rough numerical estimate of 
the spectator-scattering effect, we do not subscribe to the conclusion 
reached by these authors, that the diagrams in Figure~\ref{fig:2gluon_O8} 
are free of endpoint singularities and thus short-distance dominated. On 
the contrary, the logarithmic sensitivity to the endpoint region 
indicates that the soft spectator-scattering mechanism has one less power 
of $\alpha_s$ associated with it, and so it is formally leading with 
respect to the hard spectator-scattering contribution. 

The appearance of a leading-power soft spectator-scattering contribution 
is a novel feature of our analysis, which is specific to the case of a 
light flavor-singlet final-state meson. The QCD factorization formula for 
$B$ decays into two non-singlet mesons must be extended to account for 
this effect. We will now argue that factorization still holds in a 
generalized sense: the soft spectator-scattering contribution can be 
parameterized in terms of a convolution of the light-cone distribution 
amplitudes of the singlet meson with a perturbative kernel (the 
expression $C_P$ in (\ref{CPdef})), multiplied with a $B\to K$ ``form 
factor'' defined as the matrix element of a non-local operator. For the 
purpose of illustration we focus on the leading-power twist-2 
contribution, corresponding to the term proportional to 
$C_{8g}^{\rm eff}$ in (\ref{P2y}). Momentum conservation implies that in 
the endpoint region (where the momentum $q_2$ in 
Figure~\ref{fig:2gluon_O8} is soft) the gluon emitted from the weak 
vertex is still semi-hard, since 
$q_1^2=(p_P-q_2)^2\approx-2p_P\cdot q_2=O(m_b\Lambda)$. In the limit 
where $|q_1^2|\gg|q_2^2|=O(\Lambda^2)$, the leading contribution to the 
$\eta^{(\prime)} g^* g^*$ form factor is still given by (\ref{ffvalue}), 
however now one of the two factors $g_s$ is a non-perturbative coupling. 
Effectively, the soft gluon couples to a compact $(q\bar q)$ pair in a 
color-octet state, whose internal features it cannot resolve. The 
dependence of the soft spectator-scattering mechanism on the soft gluon 
momentum $q_2$ can be obtained from a weighted integral over a dual gluon 
field-strength tensor along the light-like trajectory of the 
$\eta^{(\prime)}$, using the formalism introduced in Section~4.3.2 of the 
second paper in \cite{BBNS}. By dimensional analysis, the  resulting 
non-local matrix element scales like a heavy-to-light form factor at 
large recoil, and we thus parameterize it as (we suppress the Wilson 
lines required to render the non-local matrix element gauge invariant)
\begin{equation}
   \int_{-\infty}^0\!ds\,s\,
   \langle\bar K|\bar s(0)\,[\nslash,\gamma^\mu]
   (1+\gamma_5)\,g_s n^\alpha {\widetilde G}_{\mu\alpha}(s n)\,b(0)
   |\bar B\rangle
   = \frac{m_B^2}{m_b}\,{\cal F}_g^{B\to K}(0) \,,
\end{equation}
where $n^\mu$ is a null-vector in the direction of the $\eta^{(\prime)}$ 
meson, and the ``form factor'' ${\cal F}_g^{B\to K}(0)$ scales like 
$(\Lambda/m_b)^{3/2}$ in the heavy-quark limit. With these definitions, 
the result for the soft spectator-scattering diagram can be represented 
as a contribution to the coefficients $\alpha_3^p(P_{q,s})$ in 
(\ref{PKamp}),
\begin{equation}
   \alpha_3^p(K P_{q,s})|_{\rm soft~spec} 
   = - \frac{3\alpha_s(\mu_h)}{8\pi N_c}\,C_{8g}^{\rm eff}(\mu_h)
   \left( \int_0^1\!dx\,\frac{\phi_{q,s}(x)}{6x\bar x} + \dots \right)
   \frac{{\cal F}_g^{B\to K}(0)}{F^{B\to K}(0)} \,,
\end{equation}
where $\mu_h=\sqrt{m_b\Lambda_h}$ with $\Lambda_h=0.5$\,GeV serves as a 
typical scale for the semi-hard gluon propagator, and as above we have 
neglected a contribution from the leading-twist two-gluon distribution
amplitude of the meson $P$. This result shows explicitly how the
factorization formula must be modified to account for the effect of soft 
spectator scattering. 

A rough estimate for the form factor ${\cal F}_g^{B\to K}(0)$ can be 
obtained by regularizing the leading-power logarithmic endpoint 
divergence in (\ref{hard}) using a cutoff such that 
$\bar y>\Lambda_h/m_B$. This yields
\begin{equation}
\label{nlest}
   {\cal F}_g^{B\to K}(0)\approx \frac{24\pi C_F\alpha_s}{N_c}\,
   \frac{f_B f_K}{m_B \lambda_B} 
   \left( \ln\frac{m_B}{\Lambda_h} - 1 \right) \approx 0.3 \,,
\end{equation}
which is of about the same magnitude as the conventional form factor
$F^{B\to K}(0)$. The resulting contribution to $\alpha_3^p(P_{q,s})$ is 
then about $2\cdot 10^{-3}$, roughly half as big as the naive 
factorization contribution in (\ref{al3NF}). 

We remark here that we are unable to reproduce the large enhancement of 
the spectator-scattering effect for $K$ mesons relative to $K^*$, which 
according to \cite{Yang:2000ce} is responsible for the large $K\eta'$ 
branching fractions. In this work the difference between pseudoscalar and 
vector mesons arises when twist-3 distribution amplitudes are included. 
We find that including the term proportional to $C_{8g}^{\rm eff}$ in 
$P^p_3(y)$ into our estimate, the coefficient of the logarithm in 
(\ref{nlest}) changes from 1 to $1+r_\chi^K/6 \approx 1.15$ for kaons and 
remains unaltered for $K^*$. We suspect that the huge effect observed in 
\cite{Yang:2000ce} is related to an error in the kernel for the twist-3 
contribution, which leads to an artificial $1/\bar y\sim m_B/\Lambda$ 
enhancement of this contribution.

\subsection{Singlet weak annihilation}
\label{subsec:singletann}

The discussion of weak annihilation effects in flavor non-singlet 
non-leptonic decays has received considerable attention recently 
\cite{Beneke:2001ev,Keum:2000ph,Lu:2000em}. Annihilation effects belong 
to the class of power-suppressed corrections to the decay amplitude, 
which cannot be calculated in the factorization approach because they 
receive leading contributions from soft partons. The analysis of the 
flavor non-singlet penguin amplitude $\alpha_4^p(M_1 M_2)$ has shown that 
annihilation effects are probably not large, yet they constitute the 
largest uncertainty in the calculation of this amplitude 
\cite{Beneke:2001ev}. We now consider the corresponding effect in the 
flavor-singlet case.

\begin{figure}
\epsfxsize=10.0cm
\centerline{\epsffile{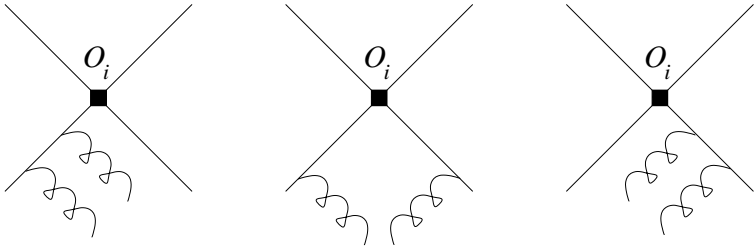}}
\vspace{0.2cm}
\centerline{\parbox{14cm}{\caption{\label{fig:annih}
Representative singlet annihilation diagrams.}}}
\end{figure}

Formally, the leading singlet annihilation amplitude is generated by the 
emission of two gluons, which form the $\eta^{(\prime)}$ meson. 
Figure~\ref{fig:annih} shows three representatives of the twenty relevant 
Feynman diagrams. The six diagrams where both gluons couple to the 
constituents of the $\bar B$ meson are part of the two-gluon contribution 
to the $B\to \eta^{(\prime)}$ form factor used as an input in the 
factorization formula, and need not be considered further. Heavy-quark 
power counting of these diagrams shows that the leading term comes from 
the diagrams with two gluons emitted from the spectator quark, and 
confirms that this contribution is of the same order in the $1/m_b$ 
expansion as the quark contribution to the form factor. To compute the 
remaining terms in the hard-scattering approach, we define the 
leading-twist two-gluon light-cone distribution amplitude by (omitting 
again a path-ordered exponential)
\begin{equation} 
\label{gluondef}
   \langle P(q)|G_\mu^{\;\;\,\rho,A}(w)\,\tilde G_{\rho\nu}^B(z)
   |0\rangle = \frac{-\delta^{AB}}{N_c^2-1}\,(\sqrt2 f_P^q+f_P^s)\, 
   q_\mu q_\nu \int_0^1\!dx\,e^{i(x q\cdot w+\bar x q\cdot z)}\,
   \phi_g^P(x) \,,
\end{equation}
where $(z-w)^2=0$, and the particular combination $\sqrt2 f_P^q+f_P^s$ 
is proportional to the flavor-singlet decay constant. The corresponding 
momentum-space projection onto the on-shell two-gluon scattering 
amplitude (with the gluon polarization vectors $\varepsilon^*$ removed) 
reads  
\begin{equation}
   \frac12\,\frac{\delta^{AB}}{N_c^2-1}\,(\sqrt2 f_P^q+f_P^s)\,
   \epsilon_{\alpha\beta\gamma\delta}\,
   \frac{q^\gamma\,\bar n^\delta}{q\cdot\bar n}\,
   \frac{\phi_g^P(x)}{x\bar x} \,,
\end{equation}
where $q$ is the momentum of the $\eta^{(\prime)}$ meson, $\bar n$ is a 
light-like vector in the direction of the kaon, and the index $\alpha$ 
($\beta$) belongs to the gluon with momentum $x q$ ($\bar x q$). In the 
FKS scheme we set 
$\phi_g^{\eta^\prime}(x)=\phi_g^{\eta}(x)\equiv\phi_g(x)$. The first term 
in the Gegenbauer expansion of $\phi_g(x)$ reads 
\cite{Terentev:qu,Ohrndorf:uz,Shifman:dk,Baier:pm,KP02} 
\begin{equation}\label{ggas}
   \phi_g(x) = 5 B_2^g(\mu)\,x^2\bar x^2\,(x-\bar x)
   + \dots \,.
\end{equation}
The coefficient $B_2^g(\mu)$ as well as all higher Gegenbauer moments 
vanish for $\mu\to\infty$. The process $\gamma\gamma^*\to\eta^{(\prime)}$ 
can in principle constrain $B_2^g(\mu)$. In our normalization 
convention\footnote{The definition (\ref{gluondef}) implies that our 
distribution amplitude is $\frac12\,\sqrt{C_F/n_f}$ times the 
distribution amplitude defined in Appendix~A of \cite{KP02}, and 
$C_F/(2 n_f)=2/9$ times the distribution amplitude assumed in Sections~3 
and 4 of the same paper, to which the determination of $B_2^g(\mu)$ 
refers.} 
the analysis performed in \cite{KP02} gives 
$B_2^g(1\,\mbox{GeV})=2\pm 3$. Note that the second Gegenbauer moment of 
the singlet quark-antiquark amplitude is much smaller, of order $-0.1$, 
which underlines the weak sensitivity of the process to the gluon 
contribution.

The result of the computation shows that we obtain a $\Lambda/m_b$ 
correction to the singlet decay amplitude from the diagrams where one 
gluon couples to the spectator quark and the other to the constituents of 
the kaon. The remaining diagrams are further suppressed in $1/m_b$. 
However, for the graphs where both gluons are emitted from the 
constituents of the kaon,\footnote{The singlet annihilation contribution 
where both gluons are emitted from the kaon can be interpreted in terms 
of a time-like $K\eta^{(\prime)}$ form factor, evaluated at $q^2=m_B^2$.} 
this additional suppression is compensated by the ``chiral enhancement 
factor'' $r_\chi^K$. For this configuration we also find that the result 
suffers from endpoint singularities. As in the flavor non-singlet case we 
can then only give an estimate of annihilation effects, based on cutting 
off the convolution integrals at parton momenta of order $\Lambda$. To 
simplify this estimate, we neglect electroweak penguin annihilation and 
note that annihilation through the tree operators $O_{1,2}$ competes with 
large tree amplitudes and is CKM-suppressed for $b\to s$ transitions. We 
therefore concentrate on annihilation effects that contribute to the 
singlet penguin amplitude. In this approximation, these are accounted for 
by substituting 
\begin{equation}
   \alpha_3^p(K P_{q,s})\to\alpha_3^p(K P_{q,s}) + \beta_{S3} \,, 
\end{equation}
where
\begin{equation}\label{bs3}
   \beta_{S3} = \left( C_6 + \frac{C_5}{N_c} \right) 
   \frac{f_B f_K r_\chi^K}{m_B^2\,F_0^{B\to K}(0)}\,
   \frac{2\pi\alpha_s}{N_c} \int_0^1\!dy\,\frac{\phi_p^K(y)}{y\bar y}
   \int_0^1\!dx\,\phi_g(x)\,\frac{x-\bar x}{x^2\bar x^2} \,,
\end{equation}
and a small contribution proportional to $C_3/N_c$ has been 
neglected.\footnote{By dropping the contribution proportional to $C_3$ we 
eliminate the diagrams with a gluon coupling to the $B$-meson spectator 
quark, which we cannot estimate in the conventional way, since the hard 
scattering amplitude depends on both the plus and minus components of the 
spectator-quark momentum, rendering the light-cone projection onto the 
$B$ meson as defined in \cite{Grozin:1996pq,Beneke:2000wa} invalid.}
Note that $\beta_{S3}$ is formally power-suppressed by a factor 
$r_\chi^K/m_b\sim 1/m_b^2$ relative to $\alpha_3^p$, similar to the 
suppression of the (numerically) dominant penguin annihilation amplitude 
for the flavor non-singlet case. Within the approximations described 
above, our result for $\beta_{S3}$ refers to penguin annihilation into a 
pseudoscalar meson ($B\to K$) with both gluons radiated from the 
constituents of the kaon, as indicated in the last diagram in 
Figure~\ref{fig:annih}. The corresponding diagrams for annihilation into 
a vector meson ($B\to K^*$) vanish for asymptotic distribution 
amplitudes. 

To obtain a numerical estimate of $\beta_{S3}$ we insert (\ref{ggas}) for 
the two-gluon distribution function, the asymptotic twist-3 distribution 
amplitude $\phi_p^K(y)=1$ for the kaon, and parameterize the 
endpoint-divergent integrals by $X_A=\int_0^1 dy/y$ as in 
\cite{Beneke:2001ev}. This gives 
\begin{equation}\label{bs3num}
   \int_0^1\!dy\,\frac{\phi_p^K(y)}{y\bar y}
   \int_0^1\!dx\,\phi_g(x)\,\frac{x-\bar x}{x^2\bar x^2}
   = \frac{10}{3}\,X_A\, B_2^g \,.
\end{equation}
Evaluating all quantities at the scale $\mu_h=\sqrt{m_b\Lambda_h}$ with 
$\Lambda_h=0.5$\,GeV, and choosing $X_A=\ln(m_B/\Lambda_h)$, we find 
\begin{equation}
   \beta_{S3}\approx -9\cdot 10^{-4}\, B_2^g \,.
\end{equation}
The value of $\beta_{S3}$ is very uncertain. Using 
$ B_2^g=2\pm 3$, and allowing $X_A$ to vary from 0 to twice
the estimate given above, we obtain 
$-9\cdot 10^{-3}<\beta_{S3}<2\cdot 10^{-3}$, but the actual size may be 
much smaller, since the central value for the gluon Gegenbauer moment may 
not reflect the true magnitude of $ B_2^g$.

\section{Phenomenological implications}
\label{sec:pheno}

We now discuss the phenomenological consequences of our results, present
numerical results for the various branching fractions, and investigate to 
which extent the striking patterns seen in the experimental data in 
Table~\ref{tab:data} can be understood in the context of QCD 
factorization. 

The leading contributions for the $B\to K\eta^{(\prime)}$ decay 
amplitudes have already been given in (\ref{silly}) and (\ref{PKamp}). 
The corresponding result for the case where $P=\pi^0$ reads
\begin{eqnarray}\label{pi0}
   {\cal A}_p(K\pi^0) &=& m_B^2\,F_0^{B\to\pi}(0)\,\frac{f_K}{\sqrt2}
    \left\{ \delta_{pu}\,\delta_{q_s u}\,\alpha_1(\pi K)
    + \sigma_{q_s} \Big[ \alpha_4^p(\pi K)
    + \frac32\,e_{q_s} \alpha_{4,{\rm EW}}^p(\pi K) \Big] \right\}
    \nonumber\\
   &+& m_B^2\,F_0^{B\to K}(0)\,\frac{f_\pi}{\sqrt2} \left\{
    \delta_{pu}\,\alpha_2(K\pi) + \frac32\,\alpha_{3,{\rm EW}}^p(K\pi)
    \right\} ,
\end{eqnarray}
where $q_s=u$ or $d$ denotes the flavor of the $B$-meson spectator quark, 
and $\sigma_u=1$ and $\sigma_d=-1$ are sign factors. At subleading order 
in the heavy-quark expansion the decay amplitudes receive corrections 
that violate factorization. Phenomenologically the most important example 
of such effects are weak annihilation contributions. Flavor non-singlet 
annihilation effects in $B\to KP$ decays can be parameterized in terms of 
two parameters $\beta_2(M_1 M_2)$ and $\beta_3^p(M_1 M_2)$ 
\cite{Beneke:2001ev,inprep}. Their effects can be incorporated by 
replacing $\alpha_1\to\alpha_1+\beta_2$ and 
$\alpha_4^p\to\hat\alpha_4^p\equiv 
\alpha_4^p+\beta_3^p$ in the expressions for the decay 
amplitudes. In addition, the term 
$\delta_{pu}\,\delta_{q_s u}\,\beta_2(K P_s)$ must be added to the 
parenthesis multiplying $f_P^s$ in (\ref{PKamp}).

For the case where the kaon is replaced by a vector meson $K^*$, one 
makes the following replacements in the expressions for the amplitudes:
$f_K\to f_{K^*}$, $F_0^{B\to K}(0)\to A_0^{B\to K^*}(0)$, 
$\phi_K(y)\to\phi_{K^*}(y)$, and 
$r_\chi^K\to r_\chi^{K^*}=(2m_{K^*} f_{K^*}^\perp)/(m_b\,f_{K^*})$, where 
$f_{K^*}^\perp$ is the (scale-dependent) transverse decay constant of the 
vector meson. In addition, the relation of the parameter $\alpha_4^p$ to 
the quantities $a_{4,6}^p$ introduced in \cite{Beneke:2001ev} changes. We 
have $\alpha_4^p(P K^*)=a_4^p(P K^*)+r_\chi^{K^*} a_6^p(P K^*)$ in 
analogy with the relation for $\alpha_4^p(P K)$, but 
$\alpha_4^p(K^* P_s)=a_4^p(K^* P_s)-r_\chi^{P_s} a_6^p(K^* P_s)$ with a 
minus sign between the two terms in the case of $\alpha_4^p(K^* P_s)$. No 
such sign was present for $\alpha_4^p(K P_s)$. Additional changes 
regarding convolutions with twist-3 vector-meson distribution amplitudes, 
and weak annihilation effects, will be detailed in \cite{inprep}.

The predictions obtained using the QCD factorization approach depend on 
many input parameters. First there are Standard Model parameters, such as 
the elements of the CKM matrix, quark masses, and the strong coupling 
constant. Of those, our results are by far most sensitive to the 
strange-quark mass. Specifically, we use $|V_{cb}|=0.041\pm 0.002$, 
$|V_{ub}/V_{cb}|=0.09\pm 0.02$, $\gamma=(70\pm 20)^\circ$, and show 
results for $m_s=100$ and 80\,MeV. Next there are hadronic parameters 
that can, in principle, be determined from experiment, such as meson 
decay constants and transition form factors. In practice, information 
about these quantities often comes from theoretical calculations. The 
corresponding model uncertainties in the form factors have a large impact 
on our results. We use (at $q^2=0$) $F_0^{B\to\pi}=0.28\pm 0.05$, 
$F_0^{B\to K}=0.34\pm 0.05$, $A_0^{B\to K^*}=0.45\pm 0.10$, and 
${\cal F}_g^{B\to K^{(*)}}=0.3\pm 0.5$ for the ``form factors'' related 
to the soft spectator interactions discussed in Section~\ref{sec:spec}. 
Finally, we need predictions for a variety of light-cone distribution 
amplitudes, which we parameterize by the first two Gegenbauer 
coefficients in their moment expansion. Experimental information can at 
best provide indirect constraints on these parameters. Fortunately, it 
turns out that the sensitivity of our predictions to the Gegenbauer 
coefficients is small, the only exception being the first inverse moment 
of the $B$-meson distribution amplitude, for which we take 
$\lambda_B=(0.35\pm 0.15)$\,GeV. (A complete list of all input parameters 
can be found in \cite{Beneke:2001ev,inprep}.)

\begin{figure}
\epsfxsize=5.0cm
\centerline{\epsffile{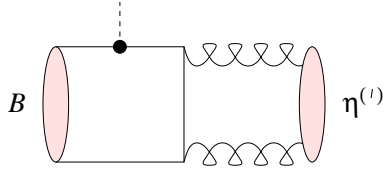}}
\vspace{0.2cm}
\centerline{\parbox{14cm}{\caption{\label{fig:ff}
Leading two-gluon contribution to the $B\to\eta^{(\prime)}$ form factor. 
The dot represents the insertion of the current.}}}
\end{figure}

An important input to our analysis are the form factors 
$F_0^{B\to\eta^{(\prime)}}(0)$ defined in terms of the 
$B\to\eta^{(\prime)}$ matrix elements of the vector current 
$\bar q\gamma^\mu b$. Unfortunately, little is known about these 
quantities. It is not justified to assume that these form factors are 
related in a simple way (modulo SU(3) violations) to the $B\to\pi$ or 
$B\to K$ transition form factors. The reason is that there is a new 
mechanism available for the case of a flavor-singlet meson, where the $B$ 
meson is annihilated by the current and the singlet meson is produced via 
the emission of two gluons. This possibility was already mentioned in 
Section~\ref{subsec:singletann}, where it was argued that the two-gluon 
emission from the light spectator quark in the $B$ meson gives a 
leading-power contribution to the $B\to\eta^{(\prime)}$ form factor. The 
corresponding diagram, shown in Figure~\ref{fig:ff}, is proportional to 
the flavor-singlet decay constant $(\sqrt2 f_P^q+f_P^s)/\sqrt{3}$. We 
thus adopt the following parameterization for the form factors ($P=\eta$ 
or $\eta'$):
\begin{equation}
   F_0^{B\to P}(0) = F_1\,\frac{f_P^q}{f_\pi}
   + F_2\,\frac{\sqrt2 f_P^q+f_P^s}{\sqrt3 f_\pi} \,,
\end{equation}
where $F_1$ and $F_2$ both scale like $(\Lambda/m_b)^{3/2}$ in the 
heavy-quark limit, and in the FKS scheme we expect that 
$F_1\approx F_0^{B\to\pi}(0)$. Given that the ratio 
$(\sqrt2 f_P^q+f_P^s)/(\sqrt 3 f_\pi)$ is about 1.2 for $P=\eta'$ and 0.2 
for $P=\eta$, we expect that the extra contribution might be significant
for the $B\to\eta'$ form factor, whereas its effect on the $B\to\eta$ 
form factor should be small. In our numerical analysis we set 
$F_1=F_0^{B\to\pi}(0)$ and take $F_2=0$ or (somewhat arbitrarily) 0.1.

Before presenting detailed numerical results for the decay amplitudes it 
is useful to consider the magnitudes of certain flavor topologies in the 
various decays. Keeping only the leading (color-allowed and not 
CKM-suppressed) electroweak penguin contributions for simplicity, we 
write
\begin{equation}\label{topologies}
\begin{aligned}
   -i {\cal A}(B^-\to K^-\eta^{(\prime)})
   &= \lambda_u^{(s)}\,(T+C+P_u+S) + \lambda_c^{(s)}\,
   (C_c+P_c+P_{\rm EW}+S) \,, \\
   -i {\cal A}(\bar B^0\to\bar K^0\eta^{(\prime)})
   &= \lambda_u^{(s)}\,(C+P_u+S) + \lambda_c^{(s)}\,(C_c+P_c+P_{\rm EW}+S)
    \,, \\
   -i {\cal A}(B^-\to K^-\pi^0)
   &= \lambda_u^{(s)}\,(T+C+P_u) + \lambda_c^{(s)}\,(P_c+P_{\rm EW})
    \,, \\
   -i {\cal A}(\bar B^0\to\bar K^0\pi^0)
   &= \lambda_u^{(s)}\,(C-P_u) + \lambda_c^{(s)}\,(-P_c+P_{\rm EW}) \,. 
\end{aligned}
\end{equation}
The tree amplitude $T$ contains terms proportional to $\alpha_1$, the
color-suppressed tree amplitude $C$ contains terms proportional to 
$\alpha_2$, and the penguin amplitudes $P_p$ contain terms proportional 
to $\alpha_4^p$ and also include the dominant non-singlet penguin 
annihilation contributions. The singlet amplitude $S$ contains terms 
proportional to $\alpha_3^p$, which within our approximations does not 
depend on the flavor label $p$. The soft spectator-scattering 
contribution as well as flavor-singlet weak annihilation are part of $S$. 
The electroweak penguin amplitude $P_{\rm EW}$ contains the terms 
proportional to $\alpha_{3,{\rm EW}}^c$, and the small contribution 
proportional to $\alpha_2(K P_c)$ is denoted by $C_c$. The various 
topological amplitudes depend on the nature of the final-state particles 
and should thus be labeled $T(KP)$ with $P=\eta,\eta',\pi^0$, etc.; 
however, in the isospin limit the amplitudes are the same for $K^-$ and 
$\bar K^0$ in the final state. A decomposition analogous to 
(\ref{topologies}) holds for the decays $B\to K^* P$. 

\begin{table}
\small
\centerline{\parbox{14cm}{\caption{\label{tab:tops}
Predictions for the dominant flavor topologies (in units of 
$10^{-9}$\,GeV). For a given final state the first line refers to 
$F_2=0$, the second to $F_2=0.1$. Errors are explained in the text.}}}
\vspace{0.1cm}
\begin{center}
\begin{tabular}{|ll|c|c|c|}
\hline\hline
\multicolumn{2}{|c|}{Mode} & 
 $e^{i\gamma}\,\lambda_u^{(s)}\,T$ & $\lambda_c^{(s)}\,P_c$
 & $\lambda_c^{(s)}\,S$ \\
\hline
 & \hspace*{-0.4cm} $F_2=0$ 
 & $4.0+0.1 i$
 & $(-65_{\,-17\,-17}^{\,+12\,+\phantom{1}9})
    +(-9_{\,-4\,-15}^{\,+4\,+15})i$
 & $(-3_{\,-13\,-2}^{\,+\phantom{1}8\,+2})+(4_{\,-5\,-2}^{\,+5\,+2})i$ \\
\raisebox{1.5ex}[-1.5ex]{$\bar K\eta'$} &  \hspace*{-0.4cm} $F_2=0.1$
 & $6.6+0.2 i$
 & $(-76_{\,-19\,-17}^{\,+13\,+\phantom{1}9})
     +(-10_{\,-5\,-15}^{\,+5\,+15})i$
 & $(-3_{\,-13\,-2}^{\,+\phantom{1}8\,+2})+(4_{\,-5\,-2}^{\,+5\,+2})i$
 \\
 & \hspace*{-0.4cm} $F_2=0$  
 & $4.9+0.1 i$
 & $(13_{\,-8\,-2}^{\,+8\,+5})+(2_{\,-1\,-4}^{\,+1\,+4})i$
 & $(-0.5_{\,-3.2\,-0.3}^{\,+2.8\,+0.3})
    +(0.6_{\,-0.7\,-0.3}^{\,+0.9\,+0.3})i$ \\
\raisebox{2ex}[-2ex]{$\bar K\eta$} &  \hspace*{-0.4cm} $F_2=0.1$
 & $5.3+0.1 i$
 & $(12_{\,-8\,-2}^{\,+8\,+5})+(2_{\,-1\,-4}^{\,+1\,+4})i$
 & $(-0.5_{\,-3.2\,-0.3}^{\,+2.8\,+0.3})
     +(0.6_{\,-0.7\,-0.3}^{\,+0.9\,+0.3})i$ \\
$\bar K\pi^0$ &  \hspace*{-0.4cm} $F_2=0$  
 & $5.8+0.2 i$
 & $(-29_{\,-8\,-8}^{\,+7\,+4})+(-3_{\,-2\,-7}^{\,+2\,+7})i$ & 0 \\
\hline
 &   \hspace*{-0.4cm} $F_2=0$ 
 & $5.4+0.1 i$
 & $(11_{\,-10\,-\phantom{1}5}^{\,+14\,+11})
    +(0_{\,-2\,-9}^{\,+2\,+9})i$
 & $(4_{\,-13\,-0}^{\,+10\,+0})+(5_{\,-8\,-0}^{\,+9\,+0})i$ \\
\raisebox{1.5ex}[-1.5ex]{$\bar K^*\eta'$} &  \hspace*{-0.4cm} $F_2=0.1$ 
 & $8.9+0.2 i$
 & $(6_{\,-10\,-\phantom{1}5}^{\,+15\,+11})
     +(0_{\,-3\,-9}^{\,+3\,+9})i$
 & $(4_{\,-13\,-0}^{\,+10\,+0})+(5_{\,-8\,-0}^{\,+9\,+0})i$ \\
 &  \hspace*{-0.4cm} $F_2=0$  
 & $6.7+0.2 i$
 & $(-28_{\,-12\,-19}^{\,+\phantom{1}8\,+\phantom{1}9})
    +(-1_{\,-2\,-16}^{\,+2\,+16})i$
 & $(0.6_{\,-2.9\,-0}^{\,+2.6\,+0})
    +(0.8_{\,-1.3\,-0}^{\,+1.4\,+0})i$ \\
\raisebox{2ex}[-2ex]{$\bar K^*\eta$} &  \hspace*{-0.4cm} $F_2=0.1$
 & $7.2+0.2 i$
 & $(-29_{\,-12\,-19}^{\,+\phantom{1}8\,+\phantom{1}9})
     +(-1_{\,-2\,-16}^{\,+2\,+16})i$
 & $(0.6_{\,-2.9\,-0}^{\,+2.6\,+0})
     +(0.8_{\,-1.3\,-0}^{\,+1.4\,+0})i$ \\
$\bar K^*\pi^0$ &   \hspace*{-0.4cm} $F_2=0$  
 & $7.9+0.2 i$
 & $(-14_{\,-3\,-10}^{\,+4\,+\phantom{1}5})
    +(-1_{\,-2\,-8}^{\,+1\,+8})i$ & 0 \\
\hline\hline
\end{tabular}
\end{center}
\end{table}

The two most important amplitudes, the penguin amplitude $P_c$ and the 
tree amplitude $T$, are given in Table~\ref{tab:tops} together with the 
flavor-singlet amplitude $S$. For the penguin and singlet amplitudes the 
second error refers to the uncertainty in the calculation of weak 
annihilation effects (more precisely, the uncertainty parameterized by 
the quantity $X_A$), and the first error contains all other parameter 
uncertainties added in quadrature, excluding an overall normalization 
uncertainty from the CKM factor $V_{cb}$. In the case of the penguin 
amplitude the parameter uncertainties are dominated by the strange-quark 
mass. In contrast, the largest uncertainty for the flavor-singlet 
amplitude comes from the quantities $\lambda_B$ and $X_H$ (defined in 
\cite{Beneke:2001ev}), which enter the calculation of the 
spectator-scattering effect. In particular, neither of the three effects 
related to the gluon content of singlet mesons, which we identified and 
calculated in this paper, constitutes a dominant uncertainty if the 
corresponding parameters lie within our estimates. The error on the tree 
amplitude is not given explicitly in the table. It is typically of order 
20\% on the real part and of order 100\% on the (small) absorptive part, 
excluding an overall normalization uncertainty from the CKM factor 
$V_{ub}$.

The table shows that with the exception of the final state 
$\bar K^*\eta'$ the flavor-singlet amplitude $S$ is not an important 
contributor to the magnitude of the decay amplitudes. In a first 
approximation the pattern of branching fractions is therefore controlled 
by the penguin amplitude $P_c$, which is indeed quite different for the 
various final states. We first note that the penguin amplitude for the 
$\bar K^*\pi^0$ final state is smaller than for $\bar K\pi^0$, because 
the $\bar K^*$ meson can be produced from a $(\bar s q)_{\rm S+P}$ 
current only through rescattering, which occurs at next-to-leading order 
in QCD factorization.\footnote{The smallness of the penguin amplitude for 
final states containing a vector meson may appear problematic in view of 
the sizeable branching fraction observed for the decay $B\to\pi^+ K^{*-}$.
The complete set of pseudoscalar--vector final states will be discussed 
in \cite{inprep}.} 
The new element for the $\eta^{(\prime)}$ final states is a contribution 
to the penguin amplitude in which the kaon (rather than the pion or 
$\eta^{(\prime)}$) picks up the spectator quark, while the 
$\eta^{(\prime)}$ is produced through the strangeness content of its 
wave function. It follows from (\ref{PKamp}) that 
\begin{equation}
\label{peneq}
   P_c(K P) \propto 
   F_0^{B\to P}(0)\,\frac{f_K}{\sqrt2}\,\hat\alpha_4^c(P K)
   + F_0^{B\to K}(0)\,f_P^s\,\hat\alpha_4^c(K P_s) \,,
\end{equation}
and the interference of the two terms determines the magnitude of the 
penguin amplitude. In the case of $\bar K\eta'$ the two terms add 
constructively to yield $-20-45=-65$ (referring to the case $F_2=0$ and 
the real parts in Table~\ref{tab:tops}). Since $f_\eta^s$ is negative 
(see Table~\ref{tab:FKS}), the second amplitude changes sign for the 
$\bar K\eta$ final state, resulting in a partial cancellation $-24+37=13$ 
of the penguin amplitude. Our calculation therefore confirms the penguin 
interference mechanism suggested in \cite{Lipkin:1990us} as the origin of 
the large $\bar K\eta'$ branching fractions. For vector mesons 
$\hat\alpha_4^c(P K^*)$ becomes smaller because the 
$(\bar s q)_{\rm S+P}$ contribution is suppressed, while 
$\hat\alpha_4^c(K^* P_s)$ becomes smaller and changes sign because the 
sign of the $(\bar s q)_{\rm S+P}$ contribution is reversed, as discussed 
at the beginning of this section. The two terms in (\ref{peneq}) 
therefore give $-9+20=11$ for $\bar K^*\eta'$ and $-11-17=-28$ for 
$\bar K^*\eta$. In cases where the penguin amplitude is suppressed by 
cancellations there is a sizeable interfering tree amplitude, which 
leads to the possibility of large direct CP asymmetries.

\begin{table}[t]
\centerline{\parbox{14cm}{\caption{\label{tab:predictions}
Predictions for the CP-averaged branching ratios (in units of $10^{-6}$), 
assuming $\gamma=70^\circ$, $|V_{cb}|=0.041$ and $|V_{ub}/V_{cb}|=0.09$. 
The first error is due to parameter variations, while the second one 
shows our estimate of the uncertainty due to weak annihilation (see text 
for explanation). The column labeled ``default'' refers to 
$m_s=100$\,MeV and $F_2=0$.}}}
\vspace{0.1cm}
\begin{center}
\begin{tabular}{|l|c|c|c|c|}
\hline\hline
Mode & Default & $m_s=80$\,MeV & $F_2=0.1$ & Experiment \\
\hline
$B^-\to K^-\eta'$ & $42_{\,-12\,-11}^{\,+16\,+27}$
 & $59_{\,-16\,-17}^{\,+22\,+41}$
 & $56_{\,-14\,-13}^{\,+19\,+31}$
 & $72.2\pm 5.3$ \\
$\bar B^0\to\bar K^0\eta'$ & $41_{\,-11\,-11}^{\,+15\,+26}$
 & $57_{\,-15\,-16}^{\,+21\,+39}$
 & $56_{\,-13\,-13}^{\,+18\,+30}$
& $54.8\pm 10.1$ \\
$B^-\to K^-\eta$ & $1.7_{\,-1.5\,-0.5}^{\,+2.0\,+1.3}$
 & $2.2_{\,-2.0\,-0.8}^{\,+2.7\,+1.9}$
 & $1.4_{\,-1.2\,-0.5}^{\,+1.8\,+1.1}$
 & $<6.9$ \\
$\bar B^0\to\bar K^0\eta$ & $1.0_{\,-1.2\,-0.4}^{\,+1.7\,+1.1}$
 & $1.4_{\,-1.7\,-0.6}^{\,+2.4\,+1.6}$
 & $0.7_{\,-0.9\,-0.4}^{\,+1.5\,+0.9}$
 & $<9.3$ \\
$B^-\to K^-\pi^0$ & $9.4_{\,-2.9\,-2.4}^{\,+3.2\,+5.6}$
 & $12.6_{\,-3.8\,-3.5}^{\,+4.3+8.2}$
 & $9.4_{\,-2.9\,-2.4}^{\,+3.2\,+5.6}$
 & $12.7\pm 1.2$ \\
$\bar B^0\to\bar K^0\pi^0$ & $5.9_{\,-2.3\,-1.9}^{\,+2.7\,+4.5}$
 & $8.5_{\,-3.1\,-2.8}^{\,+3.7\,+6.8}$
 & $5.9_{\,-2.3\,-1.9}^{\,+2.7\,+4.5}$
 & $10.2\pm 1.5$ \\
\hline
$B^-\to K^{*-}\eta'$ & $3.5_{\,-3.7\,-1.7}^{\,+4.4\,+4.7}$
 & $7.7_{\,-6.7\,-3.2}^{\,+7.6\,+8.0}$
 & $2.7_{\,-2.6\,-1.3}^{\,+3.5\,+3.9}$
 & $<35$ \\
$\bar B^0\to\bar K^{*0}\eta'$ & $2.5_{\,-3.1\,-1.5}^{\,+3.8\,+4.3}$
 & $6.3_{\,-5.8\,-2.9}^{\,+6.8\,+7.4}$
 & $1.2_{\,-1.8\,-0.9}^{\,+2.7\,+3.2}$
 & $<13$ \\
$B^-\to K^{*-}\eta$ & $8.6_{\,-2.6\,-\phantom{1}4.4}^{\,+3.0\,+14.0}$
 & $13.8_{\,-4.2\,-\phantom{1}6.7}^{\,+4.8\,+19.8}$
 & $9.1_{\,-2.7\,-\phantom{1}4.6}^{\,+3.1\,+14.3}$
 & $26.5\pm 6.1$ \\
$\bar B^0\to\bar K^{*0}\eta$
 & $8.7_{\,-2.6\,-\phantom{1}4.5}^{\,+2.9\,+14.0}$
 & $13.9_{\,-4.1\,-\phantom{1}6.7}^{\,+4.6\,+19.5}$
 & $9.2_{\,-2.7\,-\phantom{1}4.7}^{\,+3.0\,+14.2}$
 & $16.4\pm 3.0$ \\
$B^-\to K^{*-}\pi^0$ & $3.2_{\,-1.1\,-1.3}^{\,+1.2\,+4.0}$
 & $3.3_{\,-1.2\,-1.5}^{\,+1.3\,+4.8}$
 & $3.2_{\,-1.1\,-1.3}^{\,+1.2\,+4.0}$
 & $<31$ \\
$\bar B^0\to\bar K^{*0}\pi^0$ & $0.7_{\,-0.5\,-0.6}^{\,+0.6\,+2.4}$
 & $0.7_{\,-0.5\,-0.6}^{\,+0.6\,+3.0}$
 & $0.7_{\,-0.5\,-0.6}^{\,+0.6\,+2.4}$
 & $<3.6$ \\
\hline\hline
\end{tabular}
\end{center}
\end{table}

\begin{table}[t]
\centerline{\parbox{14cm}{\caption{\label{tab:predictionsACP}
Predictions for the direct CP asymmetries (in percent), assuming 
$\gamma=70^\circ$ and $|V_{ub}/V_{cb}|=0.09$, with error estimates as in 
Table~\ref{tab:predictions}. Experimental results are from 
\cite{BaBar,BelleACP}.}}}
\vspace{0.1cm}
\begin{center}
\begin{tabular}{|l|c|c|c|c|}
\hline\hline
Mode & Default & $m_s=80$\,MeV & $F_2=0.1$ & Experiment \\
\hline
$B^-\to K^-\eta'$ & $2_{\,-1\,-4}^{\,+1\,+4}$
 & $2_{\,-1\,-3}^{\,+1\,+3}$
 & $3_{\,-1\,-4}^{\,+1\,+4}$
 & $-4\pm 6$ \\
$\bar B^0\to\bar K^0\eta'$ & $2_{\,-1\,-1}^{\,+1\,+1}$
 & $2_{\,-1\,-1}^{\,+1\,+1}$
 & $2_{\,-1\,-1}^{\,+1\,+0}$
 & \\
$B^-\to K^-\eta$ & $-22_{\,-23\,-20}^{\,+16\,+24}$
 & $-18_{\,-22\,-19}^{\,+15\,+23}$
 & $-26_{\,-28\,-25}^{\,+21\,+32}$
 & \\
$\bar B^0\to\bar K^0\eta$ & $-10_{\,-17\,-5}^{\,+\phantom{1}9\,+6}$
 & $-8_{\,-14\,-5}^{\,+\phantom{1}8\,+6}$
 & $-13_{\,-31\,-8}^{\,+12\,+9}$
 & \\
$B^-\to K^-\pi^0$ & $7_{\,-3\,-10}^{\,+3\,+\phantom{1}9}$
 & $7_{\,-3\,-9}^{\,+3\,+9}$
 & $7_{\,-3\,-10}^{\,+3\,+\phantom{1}9}$
 & $-10\pm 8$ \\
$\bar B^0\to\bar K^0\pi^0$ & $-4_{\,-3\,-2}^{\,+3\,+2}$
 & $-3_{\,-3\,-2}^{\,+3\,+2}$
 & $-4_{\,-3\,-2}^{\,+3\,+2}$
 & $3\pm 37$ \\
\hline
$B^-\to K^{*-}\eta'$ & $-18_{\,-36\,-25}^{\,+25\,+30}$
 & $-10_{\,-15\,-14}^{\,+12\,+16}$
 & $-34_{\,-57\,-41}^{\,+48\,+60}$
 & \\
$\bar B^0\to\bar K^{*0}\eta'$ & $-7_{\,-15\,-4}^{\,+\phantom{1}8\,+4}$
 & $-4_{\,-6\,-2}^{\,+4\,+2}$
 & $-12_{\,-39\,-12}^{\,+15\,+\phantom{1}9}$
 & \\
$B^-\to K^{*-}\eta$ & $3_{\,-3\,-24}^{\,+3\,+25}$
 & $4_{\,-2\,-17}^{\,+2\,+17}$
 & $3_{\,-4\,-24}^{\,+3\,+25}$
 & $-5\pm 28$ \\
$\bar B^0\to\bar K^{*0}\eta$
 & $4_{\,-3\,-3}^{\,+3\,+3}$
 & $3_{\,-2\,-2}^{\,+2\,+2}$
 & $4_{\,-3\,-3}^{\,+3\,+3}$
 & $17\pm 26$ \\
$B^-\to K^{*-}\pi^0$ & $9_{\,-10\,-41}^{\,+10\,+38}$
 & $8_{\,-9\,-44}^{\,+9\,+42}$
 & $9_{\,-10\,-41}^{\,+10\,+38}$
 & \\
$\bar B^0\to\bar K^{*0}\pi^0$ & $-13_{\,-16\,-28}^{\,+15\,+24}$
 & $-12_{\,-15\,-31}^{\,+14\,+28}$
 & $-13_{\,-16\,-28}^{\,+15\,+24}$
 & \\
\hline\hline
\end{tabular}
\end{center}
\end{table}

Our predictions for the various CP-averaged branching fractions and 
direct CP asymmetries\footnote{The sign convention for the CP asymmetry 
is, contrary to \cite{Beneke:2001ev},
\[
   A_{\rm CP}(f)
   = \frac{\mbox{Br}(\bar B\to f)-\mbox{Br}(B\to \bar f)}
          {\mbox{Br}(\bar B\to f)+\mbox{Br}(B\to \bar f)} \,. 
\]} 
are collected in Tables~\ref{tab:predictions} and 
\ref{tab:predictionsACP}. Since their variations under changes of the 
CKM parameters $|V_{cb}|$, $|V_{ub}|$ and $\gamma$ within their 
respective error ranges are small compared to other uncertainties, we 
have fixed these parameters to the central values specified above rather 
than including them into the error estimates. As in Table~\ref{tab:tops},
the second error comes from weak annihilation and the first error from 
other hadronic parameters. In contrast to Table~\ref{tab:tops}, however, 
we have made the dependence on the strange quark mass explicit by showing 
results for $m_s=100\,\mbox{MeV}$ (default) and $m_s=80\,\mbox{MeV}$, and 
do not include $m_s$ into the hadronic parameter error. The following 
observations can be made:

\begin{itemize}
\item[i)]
There is a strong enhancement of some branching fractions in QCD 
factorization at next-to-leading order compared to naive factorization 
(equivalent to QCD factorization at leading order). For instance, we find 
$\mbox{Br}(K^-\eta')_{\rm LO}=12_{-5}^{+4}$ compared to 
$\mbox{Br}(K^-\eta')_{\rm NLO}=42_{\,-12\,-11}^{\,+16\,+27}$ (for default 
parameters) and $\mbox{Br}(K^{*-}\eta)_{\rm LO}=4_{-1}^{+1}$ compared to 
$\mbox{Br}(K^{*-}\eta)_{\rm NLO}
=8.6_{\,-2.6\,-\phantom{1}4.4}^{\,+3.0\,+14.0}$ (all in units of 
$10^{-6}$).
\item[ii)] 
While the calculation reproduces well the qualitative pattern of the 
known branching fractions, we find large uncertainties, principally from 
weak annihilation and the error on the strange quark mass, but also due 
to the unknown gluon contribution to the $B\to\eta^{(\prime)}$ form 
factors, parameterized by $F_2$. A smaller value of $m_s$ helps to bring 
the calculation into better agreement with the data. Where cancellations 
occur in the penguin amplitude, the uncertainty from weak annihilation is 
amplified. In view of this the significance of the $B^-\to K^{*-}\eta$ 
branching fraction, which comes out too small with default parameters, 
remains unclear. 
\item[iii)] 
Given that the $K^-\eta'$ and $\bar K^0\eta'$ final states differ only by 
a CKM-suppressed tree amplitude, we are unable to explain significantly 
different branching fractions for these two modes, if 
$\gamma=(70\pm 20)^\circ$. We therefore suspect that the apparently 
different experimental results should converge. Similar remarks apply to 
the $K^{*-}\eta$ and $\bar K^{*0}\eta$ final states. 
\item[iv)]
Even though direct CP asymmetries are generically predicted to be small 
in QCD factorization (since the strong-interaction phases of the 
amplitudes vanish in the heavy-quark limit), appreciable asymmetries can 
arise in cases where there is destructive interference between the 
dominant amplitudes. We indeed find potentially large direct CP 
asymmetries for all those modes in Table~\ref{tab:predictions} whose 
branching fraction is of order few times $10^{-6}$ or less. The 
uncertainties in these predictions are very large, typically at least a 
factor of 2.
\end{itemize}

\begin{table}
\centerline{\parbox{14cm}{\caption{\label{tab:ratios}
Predictions for some ratios of CP-averaged branching fraction, including 
error estimates for the variation of all input parameters (see text for 
explanation).}}}
\vspace{0.1cm}
\begin{center}
\begin{tabular}{|l|c|c|c|}
\hline\hline
Ratio & $F_2=0$ & $F_2=0.1$ & Experiment \\
\hline
$K^-\eta'/K^-\pi^0$
 & $4.4_{\,-1.4\,-0.2\,-0.3\,-0.4}^{\,+2.2\,+0.2\,+0.3\,+0.5}$
 & $6.0_{\,-1.8\,-0.2\,-0.3\,-0.5}^{\,+2.7\,+0.2\,+0.4\,+0.5}$
 & $5.7\pm 0.7$ \\
$K^-\pi^0/K^-\eta$
 & $5.6_{\,-\phantom{1}4.7\,-0.6\,-0.1\,-1.7}^{\,+10.0\,+0.7\,+0.1\,+2.8}$
 & $7.0_{\,-\phantom{1}6.1\,-1.0\,-0.3\,-2.2}^{\,+12.7\,+1.0\,+0.4\,+3.7}$
 & $>1.5$ \\
$\bar K^0\eta'/K^-\eta'$
 & $0.98_{\,-0.02\,-0.01\,-0.01\,-0.04}^{\,+0.01\,+0.01\,+0.00\,+0.04}$
 & $1.00_{\,-0.02\,-0.01\,-0.01\,-0.06}^{\,+0.01\,+0.01\,+0.01\,+0.05}$
 & $0.76\pm 0.15$ \\
\hline
$K^{*-}\eta'/K^{*-}\pi^0$
 & $1.1_{\,-1.1\,-0.2\,-0.6\,-0.4}^{\,+1.3\,+0.3\,+1.8\,+0.7}$
 & $0.8_{\,-0.8\,-0.1\,-0.4\,-0.4}^{\,+1.0\,+0.3\,+1.5\,+0.6}$
 & \\
$K^{*-}\eta/K^{*-}\pi^0$
 & $2.7_{\,-0.8\,-0.5\,-1.0\,-0.4}^{\,+1.1\,+0.5\,+2.0\,+0.6}$
 & $2.8_{\,-0.8\,-0.5\,-1.0\,-0.4}^{\,+1.1\,+0.4\,+2.1\,+0.6}$
 & $>0.5$ \\
$\bar K^{*0}\eta/K^{*-}\eta$
 & $1.01_{\,-0.05\,-0.03\,-0.01\,-0.14}^{\,+0.04\,+0.00\,+0.00\,+0.17}$
 & $1.02_{\,-0.05\,-0.03\,-0.01\,-0.15}^{\,+0.04\,+0.00\,+0.00\,+0.18}$
 & $0.62\pm 0.18$ \\
\hline\hline
\end{tabular}
\end{center}
\end{table}

Some of the dominant theoretical uncertainties affecting our predictions
cancel in ratios of branching fractions. The most interesting ratios are
collected in Table~\ref{tab:ratios} and compared with the experimental
data. For these predictions we perform a complete error analysis, quoting
(in this order) the uncertainties due to input parameter variations 
(other than $m_s$), weak annihilation, the strange-quark mass, and CKM 
parameters. We find that the ratios are much less sensitive to weak
annihilation and the uncertainty in the strange-quark mass than the
individual branching fractions, facilitating the comparison with 
experiment. In some cases there are, however, still sizeable 
uncertainties due to other input parameter variations, especially those 
of the heavy-to-light form factors.

A more detailed verification of the underlying mechanism of penguin 
interference will become possible once the branching fractions for which 
currently only upper limits exist have been measured. We note that we 
have also computed the corresponding branching fractions for $\Delta S=0$ 
decays, in which the final state contains a pion instead of a kaon and a 
$\rho$ meson instead of $K^*$. We find that these decays are 
tree-dominated. As a consequence the cancellations among penguin 
amplitudes are of lesser importance for the overall branching fractions, 
and no drastic signatures as in the case of strangeness-changing decays 
emerge.

\section{Summary}

Motivated by the observation of the large branching fractions for the 
decays $\bar B\to\bar K\eta'$ and the distinctive pattern of other decay 
modes with $\eta$ or $\eta'$ mesons in the final state, we have computed 
the flavor-singlet decay amplitude in rare hadronic $B$ decays using the 
framework of QCD factorization. We have considered three effects that are 
specific to singlet mesons due to their gluon content, and which occur at 
leading power in the heavy-quark expansion: the $b\to s gg$ amplitude, 
equivalent at leading order to an ``intrinsic charm'' decay constant; 
spectator scattering involving two gluons; and weak annihilation, where 
the leading-power contribution can be interpreted as a gluon contribution 
to the $B\to\eta^{(\prime)}$ form factors, and the remainder can be 
estimated in terms of the leading-twist two-gluon light-cone distribution 
amplitude of the $\eta^{(\prime)}$. A conceptually new result is that the 
spectator scattering is soft at leading order in the heavy-quark 
expansion, and hence appears to break factorization. Nevertheless, the 
long-distance contributions can be parameterized by a non-local 
$B\to K^{(*)}$ form factor, which is insensitive to the details of the 
singlet-meson wave function. In this sense factorization still holds; 
however, the factorization formula has to be amended for singlet-meson 
final states by an additional term involving this non-local form factor. 

Our numerical results indicate that the flavor-singlet decay amplitude is 
not a key factor in explaining the pattern of the observed branching 
fractions, at least if the parameters that enter this amplitude (such as 
the non-local form factor mentioned above) do not exceed substantially 
our estimates. Rather we find (as has been suggested qualitatively in 
\cite{Lipkin:1990us}) that the constructive or destructive interference 
of penguin amplitudes, where either the $K^{(*)}$ meson or the 
$\eta^{(\prime)}$ meson picks up the spectator quark, is responsible for 
this pattern. The structure of these cancellations is visible already in 
the naive factorization approach \cite{Ali:1997ex}. The important 
improvement from QCD factorization comes from the possibility to compute 
radiative corrections to the penguin amplitude, which turn out to be 
large enough to bring the predicted branching fractions into reasonable 
agreement with the data, at least for certain choices of the input 
parameters. The hadronic parameter uncertainties remain however large. We 
may therefore conclude that while it appears unlikely that one can obtain 
an accurate description of final states with singlet mesons from first 
principles, the results of our analysis clearly support the relevance of 
factorization to this class of charmless hadronic decays.

\vspace{0.3cm}  \noindent
{\it Acknowledgments:\/} 
We are grateful to Thorsten Feldmann for important comments on the 
$B\to\eta^{(\prime)}$ form factors and for careful reading of the 
manuscript, and to Peter Kroll and Kornelija Passek-Kumeri\v{c}ki for 
communicating details on their work on the two-gluon distribution 
amplitude prior to publication. The research of M.B.\ is supported in 
part by the Bundesministerium f\"ur Bildung und Forschung, Project 
05~HT1PAB/2. The research of M.N.\ is supported by the National Science 
Foundation under Grant PHY-0098631.


\end{document}